\documentclass[12pt]{article}

\usepackage{epsfig}

\usepackage{amssymb}
\usepackage{amsfonts}

\usepackage{color}
 
\def\be{\begin{equation}}
\def\ee{\end{equation}}
\def\ba{\begin{array}{c}}
\def\ea{\end{array}}

\def\ben{$$}
\def\een{$$}

\newcommand{\bea}{\begin{eqnarray}}
\newcommand{\eea}{\end{eqnarray}}

\newcommand{\kt}{\rangle}

\begin{document}


\begin{center}

{\Large \bf

Bose-Einstein condensation processes with nontrivial geometric
multiplicites realized via ${\cal PT}-$symmetric and exactly
solvable linear-Bose-Hubbard building blocks

}

\vspace{0.8cm}

  {\bf Miloslav Znojil}

\vspace{0.2cm}

The Czech Academy of Sciences, Nuclear Physics Institute,

 Hlavn\'{\i} 130,
250 68 \v{R}e\v{z}, Czech Republic

\vspace{0.2cm}

 and

\vspace{0.2cm}

Department of Physics, Faculty of Science, University of Hradec
Kr\'{a}lov\'{e},

Rokitansk\'{e}ho 62, 50003 Hradec Kr\'{a}lov\'{e},
 Czech Republic

\vspace{0.2cm}

{e-mail: znojil@ujf.cas.cz}

\end{center}

\newpage

\section*{Abstract}

It is well known that using the conventional non-Hermitian but
${\cal PT}-$symmetric Bose-Hubbard Hamiltonian with real spectrum
one can realize the Bose-Einstein condensation (BEC) process in an
exceptional-point limit of order $N$. Such an exactly solvable
simulation of the BEC-type phase transition is, unfortunately,
incomplete because the standard version of the model only offers an
extreme form of the limit characterized by a minimal geometric
multiplicity $K=1$. In our paper we describe a rescaled and
partitioned direct-sum modification of the linear version of the
Bose-Hubbard model which remains exactly solvable while admitting
any value of $K\geq 1$. It offers a complete menu of benchmark
models numbered by a specific combinatorial scheme. In this manner,
an exhaustive classification of the general BEC patterns with any
geometric multiplicity is obtained and realized in terms of an
exactly solvable generalized Bose-Hubbard model.

\newpage

\subsection*{Keywords}

.

Bose-Hubbard system of bosons;

Bose-Einstein process of condensation;

Kato's exceptional points of higher orders;

geometric multiplicity index;

classification scheme;

exactly solvable generalized Bose-Hubbard Hamiltonians;

\subsection*{Acknowledgments}

The author acknowledges the support from P\v{r}F UHK.

\newpage

\section{Introduction}

The conceptual appeal of ${\cal PT}-$symmetry (i.e., of the parity
times time reversal symmetry) currently influences several different
areas of theoretical and/or experimental physics
\cite{Christodoulides,Carlbook} as well as of the related
mathematical physics \cite{book,fre}. The birth of the concept dates
back to the mathematics of perturbation theory
\cite{Caliceti,BG,Andrianov,BM} but the real start of popularity was
only inspired by the Bender's and Boettcher's 1998 conjecture
\cite{BB} that ${\cal PT}-$symmetry of a Hamiltonian $H$ could play
a key role in a ``non-Hermitian'' formulation of quantum mechanics
of bound states (cf., e.g., reviews \cite{Geyer,Carl,ali}).

One of the basic innovations connected with the use of non-Hermitian
but ${\cal PT}-$symmetric Hamiltonians with real bound-state spectra
has been found to lie in a remarkable and highly exciting
possibility of a direct and experimentally controllable access to
the dynamical regime of a spontaneous breakdown of the symmetry.
Bender with Boettcher \cite{BB} were probably the first who managed
to simulate this process (leading to a quantum phase transition,
i.e., to an abrupt loss of the observability of the energy) using
various elementary one-dimensional single-particle local potentials.
This proved inspiring and influenced the model building efforts in
multiple areas of realistic phenomenological considerations. Among
others, several research teams turned also attention to a family of
multiparticle ${\cal PT}-$symmetric benchmark Hamiltonians $H$
called Bose-Hubbard models (see, e.g., ~\cite{Christ,zno}).

The source of the appeal of the latter models lied in the
possibility of a sufficiently realistic simulation of a specific
quantum phase transition called Bose-Einstein condensation (BEC).
From our present point of view the most inspiring analysis has been
performed in Ref.~\cite{Uwe} where a thorough mathematical
description of the BEC phenomenon has been performed using both the
linear (i.e., solvable, mathematically less complicated) and
non-linear (i.e., fully general while just perturbative or purely
numerical) versions of the underlying standard ${\cal PT}-$symmetric
Bose-Hubbard Hamiltonian.

In both of these model-building arrangements the process of the
condensation has been attributed, in the Kato's language
\cite{Kato}, to the presence of the higher-order exceptional points.
In our present paper we intend to study the former mechanism of the
quantum phase transition in more detail, extending the scope of the
approach to certain more general dynamical scenarios characterized
by nontrivial, optional geometric multiplicities of the generic
exceptional-point degeneracies.

\section{Bose-Einstein condensation}

\subsection{${\cal PT}-$symmetric Bose-Hubbard model}

\begin{figure}[h]                    
\begin{center}                         
\epsfig{file=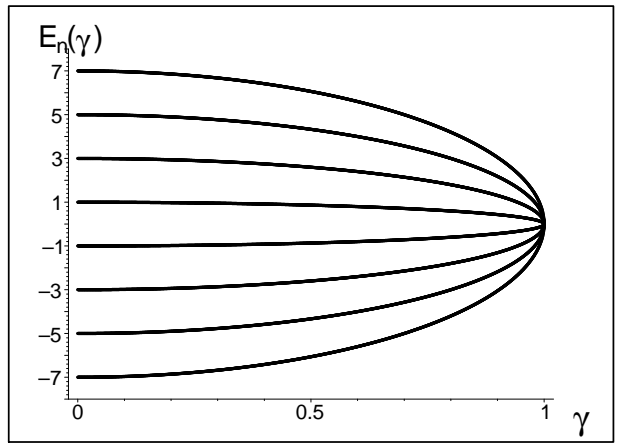,angle=270,width=0.460\textwidth}
\end{center}    
\vspace{2mm} \caption{Spectrum (\ref{rema}) for $N=8$
and its BEC degeneracy at $\gamma=1$.
 \label{lo6ja3}
 }
\end{figure}

For introduction let us
follow the
guidance provided by
Graefe et al
\cite{Uwe} who
performed a detailed
perturbation-approximation analysis
of the ${\cal PT}-$symmetric
Bose-Hubbard (BH) system living in
a small vicinity
of its BEC dynamical singularity.
One of the most elementary versions of their
non-Hermitian and
BEC-supporting family of
Hamiltonians
had the following one-parametric form
written in terms of the conventional creation and
annihilation operators,
   \begin{equation} \label{Ham1}
  \widehat{H}_{} (\gamma)=
  \left(a_1^{\dagger}a_2 + a_2^{\dagger}a_1\right) -{\rm i} \gamma
  \left(a_1^{\dagger}a_1 - a_2^{\dagger}a_2\right)\,,
  \ \ \ \ \,\gamma \in (0,1)\,.
  \end{equation}
As long as such a Hamiltonian commutes
with the operator
of the number of bosons
 \be
 \widehat{\cal N} = a^\dagger_1 a_1+ a^\dagger_2 a_2
 \label{numop}
 \ee
(with eigenvalues $N_B=1,2,\ldots$),
the
model becomes exactly solvable, with spectrum
 \be
 E_n^{}(\gamma)=n\,\sqrt{1-\gamma^2}\,,
 \ \ \ \ n \in
 {\cal S}(N)\,
 ,
  \ \ \ \ N=N_B+1
 \label{rema}
 \ee
where
 \be
 {\cal S}(N)=\{1-N, 3-N, \ldots ,N-3, N-1\}\,.
 \label{remake}
 \ee
At
$\gamma \in (0,1)$
the most
characteristic features of such a spectrum
are its reality,
equidistance and up-down symmetry:
see, for illustration, Fig.~\ref{lo6ja3} where we choose $N=8$.
In
the picture we see how
the spectrum shrinks from the $\gamma=0$
maximum \{$-7,-5,-3,-1,1,3,5,7$\}
to a complete degeneracy
at $\gamma^{} = \gamma^{(BEC)} = 1$.
A detailed derivation of formula (\ref{rema}) itself can be found in
\cite{Uwe}.

%

In the framework of our present project it
is important that
the operator version~(\ref{Ham1}) of
the unperturbed BH Hamiltonian
may be studied, most efficiently, in its
explicit infinite-dimensional
block-diagonal matrix form as derived in \cite{Uwe},
 \be
 H_{} (\gamma)=
 %
   \left (
 \begin{array}{cccc}
 H^{(2)}_{}(\gamma)&\ \ \ \ 0&0&\ldots\\
 0&H^{(3)}_{}(\gamma)&0& \\
 0&\ \ \ \ 0&H^{(4)}_{}(\gamma)&\ddots\\
 \vdots&\ \ \ \ \ &\ddots\ \ \ \ \ &\ddots
 \ea
 \right )\,.
   \label{geneve}
 \ee
Each block $H^{(N)}_{}(\gamma)$ is a tridiagonal $N$ by $N$ matrix
such that
 \be
 H^{(2)}_{}(\gamma)=
 \left[ \begin {array}{cc} -i{\it \gamma}&1
 \\{}1&i{\it
 \gamma}
 \end {array} \right]\,, \ \ \ \ \
H^{(3)}_{}(\gamma)=\left[ \begin {array}{ccc}
 -2\,i\gamma&
\sqrt{2}&0\\{}\sqrt{2}&0&
\sqrt{2}\\{}0&\sqrt{2}&2\,i\gamma\end {array}
\right]\,,
  \label{3wg}
 \ee
  \ben
 H^{(4)}_{}(\gamma) = \left [\begin {array}{cccc}
  -3\,{\rm i}\gamma &\sqrt{3}   &0  &0\\
 \sqrt{3}&-{\rm i}\gamma    &2  &0\\
  0&2  &{\rm i}\gamma  &\sqrt{3}\\
  0&0&\sqrt{3}&3\,{\rm i}\gamma
 \end {array}\right ]\,,\ \ \ \ \ \
 H^{(5)}_{}(\gamma) = \left [\begin {array}{ccccc}
  -4\,{\rm i}\gamma &2   &0  &0  &0\\
 2&-2\,{\rm i}\gamma    &\sqrt{6}  &0  &0\\
 0&\sqrt{6}    &0  &\sqrt{6}  &0\\
  0  &0&\sqrt{6}  &2\,{\rm i}\gamma  &2\\
  0&0&0&2&4\,{\rm i}\gamma
 \end {array}\right ]\,,
  \een
etc.
This representation of the Hamiltonian reconfirms the
conservation of the number of bosons $N_B=N-1$.
At every $N$, the bound-state
Schr\"{o}dinger equation degenerates
to a linear algebraic
eigenvalue problem
yielding the closed-form eigenvalues (\ref{BHepnco}).
In this
representation of the system also the
construction of the wave functions becomes
tractable non-numerically \cite{passage}.

\subsection{BEC-formation patterns}



The properties
of spectrum (\ref{rema})
are reflected by the equidistance of the
$N-$dependent
set (\ref{remake})
of the energy-level quantum numbers.
The
degeneracy of all of the levels in the BEC limit $\gamma \to 1$
is simultaneous,
 \be
 \lim_{\gamma \to 1}\,E_n^{}(\gamma)=\eta^{(N)}=0\,,
 \ \ \ \ n \in
{\cal S}(N)\,.
 \label{BHepnco}
 \ee
An analogous degeneracy also controls the
behavior of all of the eigenstates $|\psi_n^{}(N,\gamma)\kt$
of the BH operator (\ref{Ham1}),
 \be
 \lim_{\gamma \to 1}\,|\psi_n^{}(N,\gamma)\kt
 =|\chi_{(BEC)}^{(N)}\kt\,,\ \ \  \ \
 n \in S(N)\,,
 \ \ \ \ N=2,3,\ldots \,.
 \label{rekwinde}
 \ee
Precisely such a degeneracy
can be interpreted as
one of the explicit realizations
of the so called
Kato's
exceptional point of order $N$ (EPN, \cite{Kato}).

The physical
limit
$\gamma \to \gamma^{(BEC)}=1$
and the mathematical
limit
$\gamma \to \gamma^{(EPN)}=1$ appear synonymous.
Unfortunately, such a correspondence between mathematics
and physics is one-sided.
In the context of mathematics
the convergence (\ref{rekwinde}) towards a single
limiting vector $|\chi_{(BEC)}^{(N)}\kt$
must be considered
a simplification
and an
artifact of the
model.
In any sufficiently model-independent scenario
it will be necessary to replace property~(\ref{rekwinde})
(where the EPN merger has the so called geometric multiplicity $K$
equal to one)
by the more general, $K-$centered eigenvector-degeneracy
pattern with any geometric multiplicity $K \geq 1$,
 \be
 \lim_{\gamma \to \gamma^{(EPN)}}\,|\psi_{n_k}^{(EPN)}(\gamma)\kt
 =|\chi_k^{(EPN)}\kt\,,\ \ \  \ \
 n_k \in S_k^{(EPN)}\,,\ \ \
 \ \
 k=1,2,\ldots,K\,,
 \label{kwinde}
 \ee
 \be
 S_1^{(EPN)}\,\oplus \,S_2^{(EPN)}\,\oplus
 \,\ldots\,\oplus \,S_K^{(EPN)}=
 {\cal S}{(N)}\,.
 \label{diresu}
 \ee
The complete set ${\cal S}{(N)}$ of the energy quantum numbers
must be admitted to be, in general,
decomposed into
a $K-$plet of disjoint subsets.
Now, the main question to be addressed in our present paper is whether
the generalized mathematical form (\ref{kwinde}) of the EPN limit
with arbitrary $K\geq 1$ could still be
assigned a suitable physical multi-bosonic
interpretation via an exactly solvable BEC-supporting Hamiltonian, say,
of the BH type.

\section{Nontrivial geometric multiplicites $K$ at small $N$\label{sectri}}


In our recent heuristic study \cite{Borisov}
we addressed the problem on a purely methodical level.
In a trial-and-error search
for the low-dimensional toy-model matrix Hamiltonian
with nontrivial
geometric multiplicity of its degenerate EPN limit we
used the brute-force numerical methods and
performed
the search
among certain
real and more or less randomly selected matrices $H^{(6)}$.
The inspection of the parametric-dependence of the
spectra
led us to the empirical
conclusion and
conjecture of a one-to-one correspondence
between
the trivial geometric multiplicity
$K=1$
of the exceptional points
and the tridiagonality
of the matrix $H^{(6)}$ in question.

This turned out attention to
non-tridiagonal
toy models $H^{(6)}$.
The hypothesis appeared to work.
Several models supporting
the existence of exceptional points with
$K=2$ and $K=3$ were identified.
Unfortunately, our subsequent tentative
extension of the brute-force
empirical search to higher $N$
encountered severe
technical obstacles.
Our conclusions formulated in \cite{Borisov}
were, therefore, discouraging and sceptical. We argued
that beyond $N \approx 6$, the numerical EPN searches become
ill-conditioned, yielding highly unstable results
marred, in the methodical context, by the influence
of the ubiquitous rounding errors
(for details see, in particular, Appendix~A in {\it loc. cit.}).


In our present paper
we decided to accept
a different strategy.
It will be
based on a non-numerical,
strictly analytic specification of the candidates for
the suitable EPN-supporting toy-model Hamiltonians.
For this purpose
we will merely rescale
the separate fixed$-N$ components of
the $K=1$
Hamiltonian (\ref{geneve})
yielding the
trivially modified tilded forms of its submatrices, viz.,
 \be
 c_k\,H^{(2)}_{}(\gamma)=
 \left[ \begin {array}{cc} -i{\it \gamma}c_k&c_k
 \\{}c_k&i{\it
 \gamma}c_k
 \end {array} \right]=\widetilde{H^{(2)}_{(c_k)}}(\gamma)\,, \ \ \
 c_k\,
 H^{(3)}_{}(\gamma)=\left[ \begin {array}{ccc}
 -2\,i\gamma\,{c_k}&
\sqrt{2}\,{c_k}&0\\{}\sqrt{2}\,{c_k}&0&
\sqrt{2}\,{c_k}\\{}0&\sqrt{2}\,{c_k}&2\,i\gamma\,{c_k}
\end {array}
\right]=\widetilde{H^{(3)}_{(c_k)}}(\gamma)\,
  \label{c3wg}
 \ee
etc.
Obviously, any such a rescaling  with real $c_k=c_k^{(N)} \neq 0$
preserves not only the basic features of physics behind the model
but also, at any dimension $N=N_B+1$, the exact solvability
of Schr\"{o}dinger equation. At the same time,
the new, alternative, non-tridiagonal candidates for the submatrices of the
generalized $K>1$ versions of Hamiltonian (\ref{geneve})
can tentatively be constructed using suitable direct-sum combinations
of the building-blocks (\ref{c3wg}).




%

\begin{figure}[h]                    
\begin{center}                         
\epsfig{file=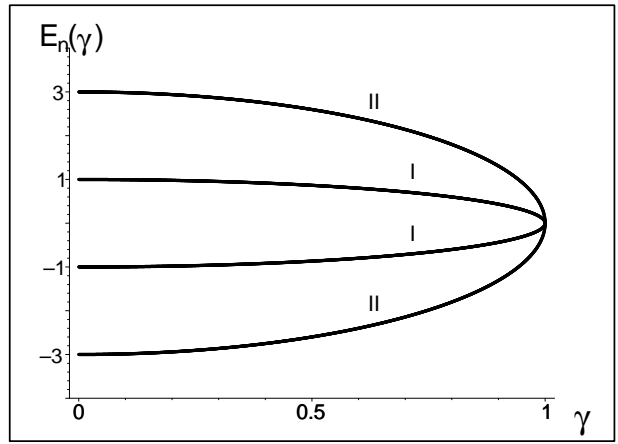,angle=270,width=0.460\textwidth}
\end{center}    
\vspace{2mm} \caption{The BEC degeneracy (\ref{rema}) at $N=4$.
 \label{loa4}
 }
\end{figure}

\subsection{Three bosons and $K=2$}
%

The first two simplest BH models with $N_B=N-1=1$ and $N_B=N-1=2$
are trivial, offering just the usual unique option with $K=1$.
Thus, our search for alternatives has to start from the next,
$N=4$ item  taken from
the $K=1$ sequence of Hamiltonians (\ref{3wg}),
with the spectrum displayed in Figure \ref{loa4}.
In the picture we marked the levels by symbols ``I'' and ``II''
in order to indicate that {\em precisely the same\,}
$\gamma-$dependent spectrum can be also
generated by an alternative Hamiltonian
defined as a
direct sum of the two $N=2$ matrices
taken out of the tilded menu (\ref{c3wg}).
This yields the innovated, $K=2$ Hamiltonian
 \be
 \label{Brasov}
    \widetilde{H^{(2)}_{(1)}}(\gamma)\oplus
   \widetilde{H^{(2)}_{(3)}}(\gamma)=
   \left [\begin {array}{cccc}
   -{\rm i}\gamma &1   &0  &0\\
 1&{\rm i}\gamma    &0  &0\\
  0&0  &-3\,{\rm i}\gamma  &3\\
  0&0&3&3\,{\rm i}\gamma
   \end {array}\right ] \sim
    \left [\begin {array}{cccc}
  -3\,{\rm i}\gamma &0   &0  &3\\
 0&-{\rm i}\gamma    &1  &0\\
  0&1  &{\rm i}\gamma  &0\\
  3&0&0&3\,{\rm i}\gamma
 \end {array}\right ]
  ={H^{(4)}_{[K=2]}}(\gamma)
  \,
 \ee
characterized by the choice of
scalings $c_1=1$ (levels ``I'')
and $c_2=3$ (levels ``II'').

Besides sharing the spectrum
(but not the eigenvectors!)
with its BH predecessor
${H^{(4)}_{[K=1]}}(\gamma)$ of Eq.~(\ref{3wg}),
the new model also illustrates several other merits of the
present philosophy.
The key point is that
the two alternative isospectral Hamiltonians can be
both made ${\cal PT}-$symmetric,
and that their imaginary parts can be kept the same (i.e., diagonal).
In this sense we can treat the right-hand-side
matrix in (\ref{Brasov})
as a canonical
{\em non-tridiagonal\,}  form of the
$K=2$ alternative to the
original $K=1$ BH building block at $N=4$.

The existence of the two  isospectral
models with different eigenvectors
may be interpreted as forcing us to
introduce a
new, {\em ad hoc\,} quantum number (say, ``color'').
Certainly, any less formal specification of such a
quantum number (or numbers)
would have to be left to the experimentalists.
Near the BEC dynamical regime only
a dedicated experiment performed over
the $N_B-$plet of bosons
will be able to distinguish between the
systems with different structures
of wave functions.
With their energy spectra fitted by
the $K-$independent formula (\ref{rema}), the
measurements would also have to extract information about
some
wave-function-dependent mean values of some other suitable
observables.


\begin{figure}[h]                    
\begin{center}                         
\epsfig{file=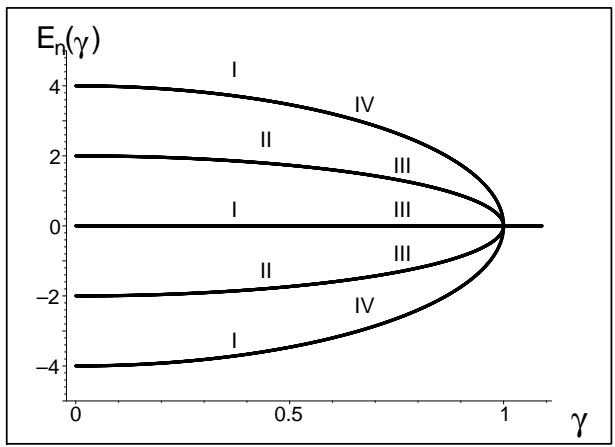,angle=270,width=0.460\textwidth}
\end{center}    
\vspace{2mm} \caption{The BEC degeneracy (\ref{rema}) at $N=5$.
 \label{loa5}
 }
\end{figure}


\subsection{Four bosons and $K=2$}

At
$N=5$ the $K=1$ Hamiltonian $H^{(5)}(\gamma)$
of sequence~(\ref{3wg})
yields
the  ``colorless''  five-level
spectrum as displayed in Figure \ref{loa5}.
The elementary combinatorics reveals that
besides such an option we can now introduce
the two alternative labelings ``I + II'' and ``III + IV''
of the states with $K=2$.
Having opted for ``I + II'' (see the picture)
we are able to reproduce the spectrum
by the direct product
 \be
  \widetilde{H^{(3)}_{(2)}}(\gamma)\oplus
   \widetilde{H^{(2)}_{(2)}}(\gamma) \sim
 \left [\begin {array}{ccccc}
 -4\,{\rm i}\gamma&0&2\,\sqrt{2}&0&0
 \\{}0& -2\,{\rm i}\gamma&0&2&0
 \\{}2\,\sqrt{2}&0&0&0&2\,\sqrt{2}
 \\{}0&2&0&2\,{\rm i}\gamma&0
 \\{}0&0&2\,\sqrt{2}&0& 4\,{\rm i}\gamma
 \end {array}\right ]=
   {H^{(5)}_{[K=2]}}(\gamma,I+II)
 \label{kc}
 \ee
(where $c_1=c_2=2$) while in the second scenario we
choose $c_1=1$ and $c_2=4$ and
have
 \be
  \widetilde{H^{(3)}_{(1)}}(\gamma)\oplus
   \widetilde{H^{(2)}_{(4)}}(\gamma) \, \sim
 \left [\begin {array}{ccccc}
 -4\,{\rm i}\gamma&0&0&0&4
 \\{}0& -2\,{\rm i}\gamma&\sqrt{2}&0&0
 \\{}0&\sqrt{2}&0&\sqrt{2}&0
 \\{}0&0&\sqrt{2}&2\,{\rm i}\gamma&0
 \\{}4&0&0&0& 4\,{\rm i}\gamma
 \end {array}\right ]=
   {H^{(5)}_{[K=2]}}(\gamma,III+IV)
 \,.
 \label{kb}
 \ee
We again decided to
arrange the basis in such a manner that the main diagonals
remain unchanged.
This renders the resulting Hamiltonian submatrices sparse, multidiagonal,
complex symmetric \cite{Garcia}
and manifestly ${\cal PT}-$symmetric.



\newpage

\section{BEC models of any dimension $N$ and multiplicity $K$}


\subsection{Canonical representation}

In the light of Eq.~(\ref{kwinde})
the non-triviality property $K>1$
of our Hamiltonian matrices
means, at any $N$, that
their canonical representation
must have,
in the EPN limit, the partitioned form
 \be
 {\cal J}^{[N]}(\eta)=
 J^{(M_1)}(\eta)\oplus
 J^{(M_2)}(\eta)\oplus
 \ldots \oplus
 J^{(M_K)}(\eta) =
   \left (
 \begin{array}{cccc}
 J^{(M_1)}(\eta)& \ 0&\ldots&0\\
 0&J^{(M_2)}(\eta)&\ddots&\vdots \\
 \vdots&\ddots &\ddots \ &0\\
 0&\ldots& \ 0&J^{(M_K)}(\eta)
 \ea
 \right )
 \,
   \label{regent}
 \ee
of a direct sum of
the standard $M_j$ by $M_j$ Jordan-block submatrices
 \be
 J^{(M_j)}(\eta)=\left (
 \begin{array}{ccccc}
 \eta&1&0&\ldots&0\\
 0&\eta&1&\ddots&\vdots\\
 0&0&\eta&\ddots&0\\
 \vdots&\ddots&\ddots&\ddots&1\\
 0&\ldots&0&0&\eta
 \ea
 \right )
 \,
   \label{JBKz}
 \ee
such that
 \be
 M_1+M_2+\ldots + M_K=N\,.
 \label{paries}
 \ee
Thus, one of the
basic characteristics of the dynamics of the system
near $\gamma^{(EPN)}$
will be an ordered version
 \be
 \pi_m(K,N) \in
 \left \{[M_1, M_2 , \ldots , M_K]
 \left | \
 M_1\geq M_2 \geq \ldots \geq M_K\geq 2\,,\ \,
   \sum_{j=1}^{K}M_j=N\,
 \right .
 \right \}\,
 \label{nupar}
 \ee
of partitioning (\ref{paries}). Up to $N=8$,
the sets of these partitionings
(having $P(N)$ elements)
are
sampled in Table \ref{iixp4}.


\begin{table}[h]
\caption{The first few sets
(\ref{nupar}) of partitions $\pi_m(K,N)$.
}\label{iixp4}

\vspace{2mm}

\centering
\begin{tabular}{||c|l|c|c||}
\hline \hline
   $N$&  $\ \ \ \ \ \ \
   \{\pi_m(K,N)\}$ & $P(N)$
     &$K_{\rm max}$\\
 \hline \hline
 \,2& \,[2] & 1&1\\
 \,3& \,[3] & 1&1\\
 \,4& \,[4],\ [2,2]  & 2&2\\
 \,5& \,[5],\ [3,2]  & 2&2\\
 \,6&  \,[6], \ [4,2],\
 [3,3],
 [2,2,2]  &  4 &3\\
 \,7&\,[7],\
 [5,2],\
 [4,3],\ [3,2,2] &  4&3\\
 \,8&\,[8],\
 [6,2],\
 [5,3],\
 [4,4],\   &&\\
 & [4,2,2],\
 [3,3,2],\
 [2,2,2,2] & 7&4\\
 \hline
 \hline
\end{tabular}
\end{table}

The trivial
partitionings
$\pi_0(1,N)=[N]$
form the leftmost column
of the Table.
The
related single, unpartitioned Jordan block (\ref{JBKz}) with $M_j$
replaced by $N$
is also most often encountered in the literature.
As the simplest
special case
of the canonical Hamiltonian
it is also the only option available
at $N=2\,$ and $N=3$. At $N>3$ the option becomes
complemented by
the simplest $K=2$ matrix~(\ref{regent})
characterized by the $m=1$ partition
$\pi_1(2,N)=[N-2,2]$, etc.
Thus,
the complete menu of the available partitions
has two elements at $N=4$ and $N=5$,
four elements at $N=6$ and $N=7$, etc.
The
growth of the number
$P(N)\,$ of the partitions
accelerates at the larger matrix dimensions $N$
(see Table~\ref{tabjed} and/or Ref.~\cite{Acc}).

In Eq.~(\ref{nupar})
the subscript
$m=0,1,\ldots, P(N)-1$
counts the partitions.
Their total number $P(N)$ is a rather quickly growing function of $N$
(see
Table~\ref{tabjed}).
Still, one can easily verify that such a subscript
(i.e., in effect, the respective partition)
cannot, by itself, serve the classification purposes.
Indeed, it it is sufficient to consider $N=6$
and to notice that there are no acceptable
direct-sum BH-type candidates for the Hamiltonian
which would be related to the partition $\pi_2(2,6)=[3,3]$.
At the same time, we will see below that
the number of the acceptable candidates
(i.e., in our notation, $a(N)$)
is equal to 6 at $N=6$ while $P(6)=4$.
Thus, some of the candidates must
necessarily share the same partition.
An additional characteristic will be required to make the classification
exhaustive and, at the same time, unambiguous.

\begin{table}[h]
\caption{The $N-$dependence of the number
$P(N)$ of partitions (\ref{nupar})
at $N\leq 20$.}\label{tabjed}
\begin{tabular}{||c||ccccccccccccccccccc||}
\hline \hline
$N$&2&3&4&5&6&7&8&9&10&11&12&13&14&15&16&17&18&19&20 \\
\hline
$P(N)$&
1&1&2&2&4&4&
7&8&12&14&21&24&34&41&55&66& 88& 105& 137\\
 \hline
 \hline
\end{tabular}
\end{table}

\subsection{Transition matrices}

The BEC- and EPN-admitting BH-type
Hamiltonians $H^{[N]}(\gamma)$ of our present interest
have to be constructed along the lines as sampled
in section \ref{sectri}
above. Once we specify their characteristic
partition (\ref{nupar})
and canonical EPN form (\ref{regent}),
we have to pay due attention also to the
Hamiltonians $H^{[N]}(\gamma)$ in the vicinity
of their EPN extreme.
A decisive technical role is then played
by the transformation yielding the
canonical representation (\ref{regent}) at $\gamma=\gamma^{(EPN)}$.
In the language of linear algebra the correspondence between
operator $H^{[N]}(\gamma^{(EPN)})$
and its partitioned canonical
Jordan-matrix representation
(\ref{regent}) is
mediated by the so called transition matrix $Q$.
This matrix, by definition,
satisfies the following
EPN substitute
 \be
 H^{[N]}(\gamma^{(EPN)})\,Q^{[\pi_m(K,N)]}
 =Q^{[\pi_m(K,N)]}\,{\cal J}^{[N]}(\eta)\,
 \label{kanon}
 \ee
for the conventional Schr\"{o}dinger equation.
The superscript $\pi_m(K,N)$
has been added here to refer to Eq.~(\ref{nupar}), i.e., to
the partition
$[M_1,M_2,\ldots,M_K]$ specifying
the direct sum ${\cal J}^{[N]}(\eta)$ of Jordan blocks
in (\ref{regent}).

In what follows we will assume that
at a fixed number of bosons $N_B=N-1$
the partitioning $\pi_m(K,N)$ is an inseparable part of
a dynamical input information
about the system even at $\gamma \neq \gamma^{(EPN)}$.
Such an input information will be carried by
the preselected
$N$ by $N$ BH-type toy-model
Hamiltonian matrix
$H^{[N]}(\gamma)$,
tractable as a small perturbation of its EPN limit
$H^{[N]}(\gamma^{(EPN)})$.
Subsequently,
in a way explained in \cite{passage}, the $\gamma-$independent
transition matrices $Q$
can be still perceived
as certain formal analogues of
unperturbed basis \cite{pert}.

In a sufficiently small EPN vicinity with $\gamma \approx \gamma^{(EPN)}$,
the Hamiltonian matrices themselves may be
assumed composed
of an
exactly diagonalizable
unperturbed component $H^{[N]}_0(\gamma)$ (equal to a
suitable BH-type direct sum of the tilded-matrix components (\ref{c3wg}))
and a perturbation.
In the EPN limit, such a perturbation must
disappear.
This means that
for our present considerations
the detailed form of this perturbation is
irrelevant. We will, therefore, ignore its influence and
drop the
zero subscript of the relevant BH Hamiltonian
as, for our present purposes, redundant,
$H^{[N]}_0(\gamma)=H^{[N]}(\gamma)$.

\subsection{Geometric multiplicities $K>1$: realization}

Our attention will be now narrowed
to the study of the BH-type systems of bosons
in which the hypothetical
available experimental information
consists of our {\it a priori\,} knowledge of
the conserved number of bosons $N_B$
and of the $\gamma-$dependence
of the spectrum as given by formula
(\ref{rema}).
Let us emphasize that
such a constraint
is well supported not only by
its occurrence in
multiple experimental setups
(where the
equidistance of the spectrum
usually reflects its ``vibrational''
character and interpretation)
but also by the significance of its purely formal merits.
In \cite{real}, for example, we showed that
even the simplest non-equidistant
choice of the square-well-type
unperturbed spectrum makes
the localization of the EPNs technically
much more complicated.

The equidistance assumption (\ref{rema})
was, initially, a rigorous result
of study of the
BH model (\ref{Ham1}).
Such a property of the system
with trivial geometric
multiplicity $K=1$ was
prescribed by formula (\ref{remake}) for indices.
In our study of the $K \geq 1$ scenarios
we now intend to proceed by analogy.
Having in mind the importance of the
exact solvability
we will complement the (entirely general)
direct-sum decomposition (\ref{diresu})
of the set of indices ${\cal S}(N)$
by a more specific, BH-motivated
additional assumption by which
all of the separate components $S_k^{(EPN)}$
will have to
share
the equidistance and symmetry
(though not the scale)
with
the global set ${\cal S}(N)$,
 \be
 S_k^{(EPN)}=\widetilde{\cal S}(M_k)
 =\{(1-M_k)c_k, (3-M_k)c_k, \ldots ,(M_k-3)c_k, (M_k-1)c_k\}
 \,,
 \ \ \ \ \ k=1,2,\ldots,K\,.
 \label{urremake}
 \ee
This means that the optional real multipliers $c_k$
will serve here as a source of an adaptive rescaling of the
subspectra.
The resulting enhanced flexibility of these
subsets of quantum numbers will be paralleled by
a rescaling (\ref{c3wg}) of the
original BH sub-Hamiltonians (\ref{3wg}). In this manner,
in a way inspired by the small$-N$ constructions of section \ref{sectri},
our general
one-parametric
$N$ by $N$ Hamiltonian matrices
will be defined as
follows,
 \be
  H^{[N]}_{}(\gamma)
  =\widetilde{H^{(M_1)}_{(c_1)}}(\gamma)\oplus
  \widetilde{H^{(M_2)}_{(c_2)}}(\gamma)\oplus \ldots \oplus
  \widetilde{H^{(M_K)}_{(c_K)}}(\gamma)
  \,.
    \label{lisNK}
  \ee
This is our ultimate ansatz.
It admits, by construction, the BEC-related
EPN singularities of arbitrary
geometric multiplicities~$K\geq 1$.
Up to $N=7$, an exhaustive list of the
mutually isospectral Hamiltonian matrices
of this type is given here in Table \ref{swexp4}.

\begin{table}[t]
\caption{Characteristics
of the isospectral
${\cal PT}-$symmetric
BH-type Hamiltonians (\ref{lisNK}) with $N \leq 7$.}\label{swexp4}
\vspace{2mm}
\centering
\begin{tabular}{||c|c|c|c|c|c||}
\hline \hline
  $N$ &  \#
     & $K$ & $m$ &  partition $\pi_m(K,N)$
     & set $\{c_1,c_2,\ldots ,c_K\}$ \\
 \hline \hline
 2&
 1& 1 & 0& [2]& \{1\}
 \\
\hline
3& 1&
  1 & 0& [3]& \{1\}
 \\
\hline 4&  1&
1 & 0& [4]& \{1\}
 \\
 & 2&
  2 & 1& [2,2]& \{1,3\}
 \\
 \hline 5&
 1& 1 & 0& [5]& \{1\}
 \\ & 2&
 2 & 1& [3,2]& \{1,4\}
 \\ & 3&
 2 & 1& [3,2]& \{2,2\}
 \\
 \hline  6&1&
 1 & 0& [6]& \{1\}
 \\ &2&
  2 & 1& [4,2]& \{1,5\}
 \\ & 3&
 3 & 3& [2,2,2]& \{1,3,5\}
 \\
 \hline
 7& 1&
  1 & 0& [7]& \{1\}
 \\  &2&
 2 & 1& [5,2]& \{1,6\}
 \\  &3&
 2 & 2& [4,3]& \{2,2\}
 \\  &4&
 3 & 3& [3,2,2]& \{1,4,6\}
 \\ &5&
  3 & 3& [3,2,2]& \{2,2,6\}
 \\ &6&
  3 & 3& [3,2,2]& \{3,2,4\}
 \\
 \hline
 \hline
\end{tabular}
\end{table}

\section{Physics behind the generalized BH model}

In the standard terminology
the two annihilation  operators
$a_1$ and
$a_2$ and the two
creation  operators
$a_1^\dagger$ and
$a_2^\dagger$ in the
initial BH Hamiltonian (\ref{Ham1})
correspond to
the two
dominant bosonic modes which may tunnel
through a hypothetical unit barrier \cite{Uwe}.
The
parameter $\gamma$ is left variable
in order to characterize the
phenomenologically most relevant
imaginary part of the on-site
bosonic-energy difference (see also
several other studies
of such a most elementary
BH bosonic system in \cite{Christ,zaUwem,[38]}).
From the point of view of physics the apparent
robustness of the $K=1$ property of such a family of
solvable BH-type models
has long been perceived as
disappointing. This seemed to imply
that the solvable picture of the BH-type
quantum
dynamics can only be
based on the conventional BH Hamiltonian (\ref{Ham1}),
not offering a description of an EPN collapse
with a non-trivial geometric multiplicity.

We have shown that an appropriate
$K>1$ generalization of the model will always
require the
direct-sum block-diagonal structure (\ref{regent})
of the canonical Hamiltonian in the EPN limit.
In this sense the highly desirable and still sufficiently realistic
BH-type realization of the general scenario
has been found to be offered
by our present model (\ref{lisNK}).
We managed to keep it exactly solvable, and
we also expect a reopening of the
questions of its predictive power
and/or tests in the laboratory.
In this direction a few immediate comments
are worth adding.

\subsection{Change of phase: two alternative physical interpretations}

In the conventional quantum mechanics of unitary
evolution
the singular values $\gamma^{(EPN)}$
were only a mathematical
curiosity. Useful in perturbation theory \cite{Kato}
but
playing hardly
any significant role in phenomenology.
Indeed, as long as
the conventional quantum theory had to be formulated in a
fixed, pre-selected Hilbert space (say, ${\cal K}$),
the basic postulate of the diagonalizability of
any meaningful quantum Hamiltonian $H(\gamma)$
appeared manifestly incompatible
with any form of
the EPN-related degeneracy
of the eigenstates.
This conclusion also sounded consistent with
the obligatory account of the unitarity and
Stone theorem \cite{Stone}.

The change of paradigm
and the turn of EPNs
into one of the fundamental physical concepts
occurred when Bender
with coauthors \cite{BM,BB}
initiated the study of certain non-Hermitian models with
relevance in the theory of relativistic
quantum fields. The
obvious phenomenological appeal of several
non-Hermitian but Hermitizable innovative forms of
interactions opened a new and promising
direction of research and discoveries
in physics \cite{Carl} as well as in mathematics
\cite{book}.
With time, these developments resulted
in an extension of the scope of the traditional
realistic models
as well as in multiple proposals of the new,
non-Hermitian but Hermitizable Hamiltonians
(see, e.g., several most recent
reviews in \cite{Christodoulides,Carlbook}).

In particular, a compatibility of the new theory with the Stone
theorem has been achieved via an {\it ad hoc\,} amendment of the
inner product
in ${\cal K}$. As a consequence, the later space
proved converted into a new, strictly physical Hilbert space
(say, ${\cal H}$) in which
the Hamiltonian re-acquired its necessary self-adjoint status \cite{ali}.
In our present paper the main consequence of the
perspective of
working with the two Hilbert spaces ${\cal K}$ and ${\cal H}$
is twofold. Firstly, the
non-Hermiticity of $H(\gamma)$
in ${\cal K}$
admitted the existence of
the EPNs at a real parameter $\gamma^{(EPN)}$.
Secondly,
the mathematical consistence of the theory
has been found achieved by the observation that ${\cal H}$
simply ceases to exist in the singular limit of
$\gamma\to \gamma^{(EPN)}$.

Formally speaking, the admissibility of the
reality of the EPNs
was a consequence of the Hermitizability
of $H(\gamma)$ or, indirectly, of its
${\cal PT}-$symmetry \cite{Carl}.
Still, it was also necessary to take into account that in general,
the physical phenomena covered by
non-Hermitian
models
may be found extremely sensitive to random
perturbations \cite{pert,Trefethen,Viola}.
For the latter reason, it is necessary
to distinguish between the implementations of the theory,
mainly in the
two entirely different experimental setups.
In one (let us call it, for our present
terminological purposes, ``approach A''),
many theoreticians
are trying to predict the qualitative features of the
dynamics
of the system {\em after\,} its perturbation.
Interested readers can find an extensive sample of
the results
of such a type in paper \cite{Uwe}.

In that paper
the authors described
several alternative unfolding scenarios of the EPN degeneracy,
i.e.,
several eligible patterns of
the complexification of the energy spectrum under
a ``strong'', unitarity-violating perturbation.
In the context of our present paper we
would slightly prefer an alternative philosophy
(let us call it  ``approach B'').
In contrast to approach A (which analyzes, basically, the
non-unitary effective open-system
dynamics),
this approach is restricted
to the
analysis of the unitary, closed-system behavior of the
quantum system of interest.
{\em Before\,} it passes through its
EPN-mediated phase transition, and {\em before\,}
the reality (i.e., the observability)
of the energy spectrum becomes lost.

One of the physics-related consequences is
the emphasis put upon the conservation-law role of
operator (\ref{numop})
which describes the number of bosons in the system. We
declared
such a number a strictly conserved quantity.
Under this assumption we are allowed to
work with a fixed and finite number of bosons $N_B$.
Also the
number of the energy levels is kept finite and equal to
$N=N_B+1$.
Last but not least, it is worth adding that
even the  inclusion of perturbations
changing the number
of bosons need not necessarily make the model
prohibitively complicated (in this respect,
interested readers could
consult, e.g., Ref.~\cite{without}).

\subsection{The role of ${\cal PT}-$symmetry}

The modern applications of quantum theory range from condensed
matter physics and materials science up to chemistry and
engineering. In this context many theoretical methods often become
popular with a certain delay. One of the good examples is the
concept of parity-times-time-reversal symmetry (${\cal
PT}-$symmetry) which was initially perceived as a curiosity
possessing just a few purely abstract mathematical
applications (say, during the studies of
some subtleties in perturbation theory -- see, e.g.,
Refs.~\cite{Caliceti,BG,Alvarez}).
Needless to add,
the phenomenological relevance of the concept of
${\cal PT}-$symmetry is now widely recognized and
quickly growing
(see, e.g., several most recent
reviews collected in books \cite{Christodoulides,Carlbook}).
At present such a tendency
still keeps being productive
in several traditional areas of physics.
It
offers new insights also in some subtle mathematical aspects of
the theory of
quantum phase transitions.

In spite of the manifest
non-Hermiticity of the ${\cal PT}-$symmetric
candidates $H$
for Hamiltonians,
these operators were shown
eligible as generators of unitary evolution
\cite{Geyer,ali}.
In this context,
one of the basic methodical assumptions accepted
in the current
literature on BH models \cite{passage,Borisov,nje4a5}
was that
the infinite-dimensional matrix (\ref{geneve})
as well as all of its
separate submatrices (\ref{3wg})
had to be complex symmetric, {\em tridiagonal\,}
and ${\cal PT}-$symmetric, with
${\cal P}$ equal to an antidiagonal unit matrix,
and with symbol
${\cal T}$ representing an antilinear operation of
Hermitian conjugation (i.e., transposition plus
complex conjugation).
This led to the conclusion (or rather conjecture) that
in the EPN limit (i.e., at the instant of
the loss of diagonalizability),
the canonical representation of
{\em every\,} $N$ by $N$ submatrix $H^{(N)}_{}(\gamma^{(EP)})$
can be given the form of the $N$ by $N$ Jordan matrix (\ref{JBKz})
with, due to ${\cal PT}-$symmetry,
$\eta=0$.

In mathematical terminology
this seemed to imply that in the BH models,
the geometric multiplicity $K$
of the EPN in question had to be,
at any preselected value of $N_B=N-1$, {\em always\,}
equal to one \cite{Borisov}.
And it is precisely this belief and scepticism which are
disproved by our direct-sum ansatz (\ref{lisNK}).

\section{Combinatorics behind the classification}

The
relevance of the BEC models with $K>1$
is twofold. First, the differences between
partitions (\ref{nupar})
may
inspire new experiments,
especially in the dynamical regime close to
the EPN singularity. Second,
the number of non-equivalent partitionings
(\ref{nupar})
as well as the  number of admissible Hamiltonians
(\ref{lisNK})
are both quickly increasing functions
of the matrix
dimension $N$.
Obviously, in the language of mathematics,
the occurrence of non-minimal
geometric multiplicities $K>1$ may be expected to
dominate, especially among the random
Hamiltonian matrices.
For both of these reasons,
the currently open problem
of the classification of the models at large $N$
would certainly deserve an enhanced attention.

\subsection{Classification scheme}

At the not too large matrix dimensions $N$, our present
condition of equidistance of the energy levels
remains reasonably restrictive. At the smallest $N$
one even encounters
not too numerous alternative isospectral BH models
with Hamiltonians (\ref{lisNK}).
Once we denote the number
of non-equivalent BEC-degeneracy scenarios
by a dedicated symbol $a(N)$, we may notice that
the growth of this value
at intermediate $N$ still remains comparatively slow,
especially at the even $N$ (see Table \ref{uxp4}).
Still, asymptotically,
the growth of the value of $a(N)$
with the growth of $N$
becomes exponentially quick. Even then, the
sequence $a(N)$ keeps exhibiting a parity-related
irregularity (see
\cite{oeisodd} and
\cite{oeiseven} for details).

\begin{table}[h]
\caption{The first few
counts $a(N)$
of the generalized BH models of Eq.~(\ref{lisNK}).}\label{uxp4}

\vspace{2mm}

\centering
\begin{tabular}{||c||cccccccccccccc||}
\hline \hline
$N$&2&3&4&5&6&7&8&9&10&11&12&13&14&15 \\
\hline
$a(N)$&1&1&2&3&3&6&4&11&6&17&7&32&8&47\\
%
 \hline
 \hline
\end{tabular}
\end{table}

For the purposes of an explicit classification of the models,
it is necessary to consider not only the (conserved)
number of bosons $N_B=N_B(N)=N-1$ and
the geometric multiplicity $K$ of the BEC degeneracy but also
the selection of partition $\pi_m(K,N)$ and of the set
of scaling parameters $\{c_1,c_2,\ldots ,c_K\}$.
All of these choices have to be made
compatible with the direct-sum decomposition
formula (\ref{lisNK}). The resulting exhaustive list of
the Hamiltonians
is
sampled in Table \ref{swexp4}.
The Table offers
a clarification of the
relationship between the scarcity of our initial
BH Hamiltonians in their matrix representations
(\ref{geneve}) or (\ref{3wg})
(where we always had $K=1$)
and the abundance of
their $K>1$ EPN-supporting generalizations
$H^{[N]}(\gamma)$.

Although the
construction of Table \ref{swexp4} was just an elementary combinatorics,
its extension to the larger dimensions $N$
is still an open question.
The construction
becomes tedious, well suited for the
computer-assisted enumeration.
Incidentally,
the suitable algorithms proved different for the
odd and even $N$.
Their most efficient
versions were proposed by Andrew Howroyd
and may be found published,
in
the on-line encyclopedia of integer sequences, under the
respective coordinates
\cite{oeisodd} and
\cite{oeiseven}.

Our  spectrum-equidistance
constraint restricts, in
Hamiltonian~(\ref{lisNK}),
the freedom of our choice of
the partitions $\pi_m(K,N)$
and of the scalings $c_k$.
From a descriptive and phenomenological point of view,
the family of the resulting generalized BH
Hamiltonians still remains sufficiently rich.
In the manner illustrated in Table \ref{swexp4}
one only has to notice that at a fixed (i.e.,
unrestricted but conserved)
value of $N$,
a straightforward classification of these
Hamiltonians is not provided
by the multiplicity $K$ (even at $N=5$, the $K=2$ option
is already shared by the two different Hamiltonians)
nor by the partitions (at $N=6$, partition $[3,3]$ is
not realized at all)
but only by their combination with the eligible multiplets
of scalings $\{c_1,c_2,\ldots ,c_K\}$.

\subsection{An alternative notation\label{64}}

A rather abstract nature of the latter criterion
indicates that
a more physics-oriented guide to the classification
could and might be sought in
a return to the
explicit spectrum (\ref{rema}).
We know that
the  unique) $K=1$ element
$H^{(N)}_{}(\gamma)$
of the BH-matrix sequence (\ref{3wg})
is unambiguously specified by the
spectrum.
In this sense
the Hamiltonian matrix becomes uniquely identified
by the $\gamma=0\,$ spectrum {\it or\,} by the symbol
${\cal S}(N)$. Thus, with $N=5$,
for example, we have ${\cal S}(5)
=\{-4,-2,0,2,4 \}$
admitting the two
centrally symmetric and equidistant
decompositions
$\{-2,0,2 \}\bigcup\{-4,4 \}$
and $\{-4,0,4 \}\bigcup \{-2,2 \}$.
Both may be most easily attributed
the respective Hamiltonian matrices (\ref{kb}) and (\ref{kc}).

The latter, more intuitive and spectrum-representing
version of notation admits an abbreviation
which has been used in Refs.~\cite{oeisodd} and
\cite{oeiseven}.
The
centrally symmetric and equidistant $K=1$ multiplets
${\cal S}(N)=\{1-N,3-N,\ldots, N-1 \}$
were represented there by an abbreviated symbol
$\{024\ldots (2J)\}$ (for odd $N=2J+1$)
or
$\{135\ldots (2J-1)\}$ (for even $N=2J$).
One also needed the auxiliary
rescaled symbols like
$p\,\{024\ldots\}=\{0(2p)(4p)\ldots\}$
and
$q\,\{135\ldots\}=\{q(3q)(5q)\ldots\}$.
In this notation, therefore, the ``colorless'' $K=1$
building-block items
forming the sequence (\ref{3wg})
are assigned the respective abbreviated
flavor-specifying
indices
$\{1\}$ (at even $N=2$),
$\{02\}$ (at odd $N=3$),
$\{13\}$ [at even $N=4$],
$\{024\}$ [at odd $N=5$], etc.

Once we further move to the simplest nontrivial ``colored''
building-block (\ref{Brasov}) with $N=4$ and $K=2$,
our new form of the reference index will simply be the union
$\{1\}\bigcup\{3\} \equiv \{1\}\{3\}$.
Similarly, at  $N=5$ and $K=2$ we will have
the index $\{02\}\bigcup\{4\} \equiv \{02\}\{4\}$
for matrix (\ref{kb}), and the second possible index
$\{04\}\bigcup\{2\} \equiv \{04\}\{2\}$ for
the other matrix (\ref{kc}).

Once we admit an arbitrary value of
dimension $N$,
our present, physics-oriented
problem of the classification
of all of the possible $N$ by $N$ sub-Hamiltonians
becomes equivalent to
the purely combinatorial problem of an exhaustive
generation of all of the possible decompositions
of index $\{024\ldots 2J\}$ (for odd $N=2J+1$)
or index
$\{135\ldots 2J-1\}$ (for even $N=2J$)
into the appropriate sub-indices.

\begin{table}[h]
\caption{The first few generalized BH models
in the alternative notation of paragraph \ref{64}
 (the trivial, tridiagonal-matrix BH items of Eq.~(\ref{3wg})
 are boxed).}\label{pexp4}

\vspace{2mm}

\centering
\begin{tabular}{||c||l||c||l||c||}
\hline \hline
   $N$&   partitions (\ref{nupar}) of $N=N_B+1$ & \#
     &  isospectral-Hamiltonian indices & \# \\
 \hline \hline
 2& {2} & 1& \fbox{\{1\}}& 1\\
 3& {3} & 1&  \fbox{\{02\}}& 1\\
 4& {4}, {2+2}  & 2&  \fbox{\{13\}}, {}{\{1\}$\!$\{3\}} & 2\\
 5& {5}, {3+2}$\,^{(a)}$  & 2& \fbox{\{024\}},
  {}{\{02\}$\!$\{4\}},
 {}{\{04\}$\!$\{2\}} & 3\\
 6& {6}, {4+2},
 {3+3}$\,^{(b)}$,
 {2+2+2}  &  4 &  \fbox{\{135\}},
 {}{\{13\}$\!$\{5\}}, {}{\{1\}$\!$\{3\}$\!$\{5\}} & 3\\
 \hline
 7&{7},\
 {5+2},\
 {4+3},\ &  4&
  \fbox{\{0246\}}, {}{\{024\}$\!$\{6\}},
  {}{\{04\}$\!$\{26\}},
 & 6\\
 &
 {3+2+2}$\,^{(c)}$ & &
 {}{\{02\}$\!$\{4\}$\!$\{6\}},
 {}{\{04\}$\!$\{2\}$\!$\{6\}},
 {}{\{06\}$\!$\{2\}$\!$\{4\}} & \\
 \hline
 8&{8},\
 {6+2},\
 {5+3}$\,^{(b)}$,\
 {4+4}$\,^{(b)}$,\  {4+2+2},\ & 7&
   \fbox{\{1357\}}, {}{\{135\}$\!$\{7\}},
  {}{\{13\}$\!$\{5\}$\!$\{7\}}, & 4\\
 &
 {3+3+2}$\,^{(b)}$,\
 {2+2+2+2} & & {}{\{1\}$\!$\{3\}$\!$\{5\}$\!$\{7\}} & \\
 \hline
 \hline
 \multicolumn{5}{||l||}{$^{(a)}$ realized twice}\\
 \multicolumn{5}{||l||}{$^{(b)}$ not realized}\\
 \multicolumn{5}{||l||}{$^{(c)}$ realized thrice}\\
 \hline
 \hline
\end{tabular}
\end{table}

The first few samples of the modified classification
may be found displayed in our last Table \ref{pexp4}.
More comments on such a notation
may be also found in
\cite{oeisodd}
and \cite{oeiseven}.
In particular,
these references offer an algorithm for the evaluation
of
the ``count of the colors''
$a(N)$ at any $N$.


\section{Discussion}

In our paper we managed to clarify
certain qualitative properties of the
BEC degeneracies.
For the sake of definiteness we had to assume
that the set of indices ${\cal S}{(N)}$
was defined by Eq.~(\ref{remake}).
Nevertheless,
only our independent requirement of the BH-type
physics
extended the same requirement to all of the components of
the direct sums
(\ref{diresu}).
Otherwise, the form of these
subsets $S_k^{(EP)}$ would
remain less constrained, numbering strictly just the
eigenstates which had to
coincide with the $k-$th ket $|\chi_k^{(EPN)}\kt$ in the limit.
Thus, every such a subset would only be required to contain
two or more quantum numbers $n$.

\subsection{The specific features of bosons}

In the light of the latter comment we believe that
the price to pay for the enhanced freedom
(consisting in the loss of the connection of physics with the
BH model) would be too high and hardly acceptable.
Indeed, the BH model mimics
the
experimentally
realizable
double-well arrangement in which, say,
the cold bosonic atoms are injected
in one well and, simultaneously,
extracted from the other one \cite{Klai,Cart}.
Moreover, the model also reflects
the key difference between fermions and bosons
because
the ``natural'' Pauli exclusion principle
only applies to the the former class of the quantum particles.
Bose with Einstein \cite{Bose}
were among the first to notice that in the systems of bosons,
this opens the possibility of a
specific quantum phase transition (which is, at present,
widely known as the
Bose-Einstein condensation \cite{Douglas}).

Both the theoretical and experimental
aspects of the BEC idea proved
particularly useful and inspiring
in the condensed matter physics.
The condensates were successfully
described there, typically, by various nonlinear
forms of Schr\"{o}dinger equation
\cite{Kostya}.
In our present paper we paid attention to the
less widespread simulation of the condensation processes
in which the underlying dynamical
equations are required
${\cal PT}-$symmetric, i.e., invariant with respect to the
simultaneous action of
parity ${\cal P}$ and of the antilinear time reversal ${\cal T}$ \cite{Carl}.
As we already indicated above,
the scope and impact of such a methodical innovation is
not yet fully explored, with the existing results
ranging from
an amendment of our understanding of the topological phase transitions
(say, in a non-Hermitian Aubry-Andre-Harper model
\cite{citkorid} or in quasicrystals \cite{kukorid})
up to the new approaches to the
perception of the conservation laws \cite{orid},
and from the theoretical studies of the
interference between channels
\cite{skost} and of the mechanisms of squeezing \cite{rekorid}
up to the detailed, experiment-oriented
simulations of the properties of the
specific BEC-type condensates
\cite{Cart}.
In this framework, promising results are also being obtained
in the area of related mathematics.

Along these lines we showed that although
the model (\ref{Ham1}) is, in some sense,
naive, its exact solvability
represents an important advantage.
In fact, precisely this property
encouraged us to propose a generalization
which
enhanced the theoretical scope
while still remaining realistic.
We may summarize that our ``benchmark'' amendment of the model
seems to offer a fairly satisfactory
picture of the physical BEC-related reality
which is obtained, in addition, by
the purely non-numerical means.

%


\subsection{The problem of the non-uniqueness of the model\label{sekcedva}}

In the above-mentioned open-system-theory approach A
(which may be found more thoroughly explained
also in monograph \cite{Nimrod}),
the one-parametric ${\cal PT}-$symmetric BH model (\ref{Ham1})
can be interpreted as one of the phenomenological descriptions of
a realistic and potentially unstable quantum system of bosons
in which their number $N_B$ itself is conserved.
In such an effective-description approach the model
simulates
the behavior of an open system
exposed, say, to the influence of an
environment
in a way characterized, globally, by the parameter.
The manifest non-Hermiticity and non-unitarity of the model
reflects, effectively, the randomness of the environment.

In our presently preferred approach B,
the physical interpretation of the model is different,
treated fully in the spirit of the original Bender's
philosophy \cite{Carl}.
This, naturally,
enhances the impact of ${\cal PT}-$symmetry, and it
changes also a part of the related mathematics.
First of all,
the conventional Hilbert space ${\cal K}$
must be declared unphysical because in this space
the evolution (controlled by Hamiltonian $H$ which is
non-Hermitian) would be,
in the light of the well known Stone theorem \cite{Stone},
non-unitary.
Incidentally, the apparent emerging paradox has a virtually
elementary resolution: Along the lines discovered and
discussed already in
older literature \cite{Geyer,Dyson},
it is sufficient to endow the same space ${\cal K}$
with an amended inner product.
Interested readers should search for the necessary
mathematical details elsewhere \cite{book,Geyer,ali}.
For introduction it is just sufficient to keep in mind
that the amendment of the inner product
converts the unphysical Hilbert space ${\cal K}$
into its
physical Hilbert-space
alternative ${\cal H}$ in which $H$ becomes, by construction,
Hermitian {\it alias\,} self-adjoint.

In approach B, the first task is to demonstrate the
existence of a suitable inner product
(see, e.,g., an extensive discussion of this
aspect of the theory in \cite{Lotor}),
the necessary condition of which is the reality of the spectrum.
This means that, in some sense, the two approaches A
and B meet
at the boundary where $\gamma^{(EPN)}=1$.
In both of the neighboring dynamical regimes, after all,
the hypothesis of the existence of the EPN boundary
opens an exciting theoretical possibility of description
of an EPN-mediated phase transition in the system.

In our study, we
had mainly in mind approach B.
We considered, exclusively,
the positive and unitarity-compatible parameters
$\gamma<\gamma^{(EPN)}$.
In such a subinterval
of parameters it is possible to guarantee that
the necessary physical
Hilbert space ${\cal H}$ does really exist
because the spectrum of $H$ is real \cite{ali}.
In such a context
there emerges one of the most interesting
theoretical questions: Would
the experimental confirmation of the
BH-model-based predictions
mean and imply that
the validity of the theoretical model is confirmed?
Obviously, it is not so because
in an inverse
BH bound-state problem,
the candidate
for the simulation of a given energy spectrum
is not unique. Constructively we showed that
a given solvable BH-type Hamiltonian
can  be complemented
by its many isospectral alternatives,
yielding even exactly the same
parameter-dependence of the energy spectrum.
This implies that in a complete experiment
one would have to measure also some wave-function-dependent
features of the system.

\subsection{A remark on the theory of quantum phase transitions}

{\em Before\,} the BEC/EPN collapse
(i.e., whenever $|\gamma|<1$),
the spectrum
of all of our present BH Hamiltonians
remains real and observable.
{\em At the instant\,} of collapse,
one can speak about a quantum phase transition \cite{denis}.
{\em After\,} the system has passed {\em through\,}
such an EPN singularity,
one has a choice between the
above-mentioned open-system evolution scenarios
of approach A
(admitting
the loss of the reality
of the spectrum)
and the closed-system scenario B in which the escape from
the EPN singularity proceeds
through a fine-tuned corridor of
unitarity
as sampled, in the BH context, in \cite{passage}.

In a broader phenomenological context,
one of the most typical aspects of the change of phase
of any quantum system
can be seen in the loss of observability
of at least one of its observable characteristics.
Among the most popular scenarios of phenomenological interest
one encounters the
loss of observability of the
energy levels,
i.e., in Schr\"{o}dinger picture \cite{Messiah},
of at least some
of the parameter-controlled eigenvalues
of the Hamiltonian.

From the point of view of mathematics,
it makes sense to prefer the
study of the Hamiltonians which are analytic functions
of the parameter, $H=H(\gamma)$.
In the generic, infinite-dimensional and purely numerical models
the loss of the observability
can be then perceived as caused,
in a partial analogy with our present
solvable models, by the
limiting transition
$\gamma \to \gamma^{(EPN)}$
in which the limiting value of the parameter is the Kato's
exceptional point of some finite order $N$.
One can say that at in such a limit
there also exists a finite subset $S=S^{(EPN)}$ of
quantum numbers $n$
for which the energies
as well as the wave functions merge.
Due to our present intention of keeping the models solvable,
it was only necessary to decompose
the discrete set $S^{(EPN)}$
in
the very special, BH-related
direct sum (\ref{diresu}).
Naturally,
the same method and approach should also work in the
descriptions of the quantum systems which
are not solvable exactly.

\section{Summary}

The authors of Ref.~\cite{Uwe} emphasized the deep methodical
relevance of model (\ref{Ham1}) in the vicinity of its BEC
degeneracy. They even developed a specific {\it ad hoc\,}
perturbation theory and described some of the consequences of the
inclusion of perturbations. Several alternative BEC-related
perturbation-influence results can be also found in
Refs.~\cite{znopert}. These studies clarified multiple quantitative
aspects of the condensation.

In our present project
we managed to construct a BH-based simulation
of the general BECs in a
solvable-model realization.
The resulting picture of
a generic EPN degeneracy appeared to have two aspects.
The positive one is that in our innovative family of the
BH-type models the EPN limit has been made
nontrivial, characterized by any kinematically
admissible geometric multiplicity $K$.
In our methodically oriented considerations,
on the other hand, our results only involved
the study of the Hamiltonian.
Our description of the system of bosons
did not involve
any other quantities which would be
measurable and which would
enable us to clarify the differences between our
non-equivalent BH-type models. Thus,
the clarification of these differences
(characterized here just by a rather formal
proposal
of an auxiliary quantum number)
will certainly be a task
for some forthcoming studies of the BH-type systems of bosons.

From the mathematical point of view the
most exciting aspect of our generalized
BH models of BEC was, certainly, the
discovery of the feasibility of the
enumerative
classification of all of the admissible
direct-sum Hamiltonians (\ref{lisNK}).
This was an interesting
combinatorial task which found the necessary
background in the specialized literature.
In such a formulation
one cannot say that the task
is already completed.
Indeed, the two Howroyd's
brute-force algorithms of the
enumeration
of all of the phenomenologically meaningful
unions (\ref{diresu}) of the
specific equidistant and symmetric
spectral subsets (\ref{urremake})
of the energy quantum numbers
would certainly deserve to be complemented by
some less implicit
(e.g., recurrent) alternatives
(provided only that they do exist at all).


\end{document}